\documentclass[12pt,preprint]{aastex}

\title{Properties of Disks and Bulges of Spiral and Lenticular Galaxies 
in the Sloan Digital Sky Survey}
\author{N. Oohama$^{(a)}$, S. Okamura$^{(b)}$, M. Fukugita$^{(c,d)}$, N. 
Yasuda$^{(d)}$, O. Nakamura$^{(e)}$}
\affil{$^{(a)}$Department of Astronomy, University of Tokyo, Hongo,
Tokyo 113-0033, Japan\footnote{present address: Hitachi Software
Engineering, Co. Ltd.}}
\affil{$^{(b)}$Department of Astronomy and Research Center for the Early 
Universe, University of Tokyo, Hongo, Tokyo 113-0033, Japan}
\affil{$^{(c)}$Institute for Advanced Study, Princeton, NJ 08540 USA}
\affil{$^{(d)}$Institute for for Cosmic Ray Research, University of 
Tokyo,\\Kashiwa 277-8582, Japan}
\affil{$^{(e)}$Graduate School of Political Science,
Waseda University, \\Shinjuku, Tokyo 169-8050, Japan}

\begin{document}

\maketitle
\section*{Abstract}
A bulge-disk decomposition is made for 737 spiral and lenticular
galaxies drawn from a SDSS galaxy sample for which morphological types
are estimated.  We carry out the bulge-disk decomposition using the
growth curve fitting method.  It is found that bulge properties,
effective radius, effective surface brightness, and also absolute
magnitude, change systematically with the morphological sequence; from
early to late types, the size becomes somewhat larger, and surface
brightness and luminosity fainter. In contrast disks are nearly
universal, their properties remaining similar among disk galaxies
irrespective of detailed morphologies from S0 to Sc. 
While these tendencies were often discussed in previous studies, 
the present study confirms them based on a large homogeneous 
magnitude-limited field galaxy sample with morphological types 
estimated. The systematic change of bulge-to-total luminosity ratio, 
$B/T$, along the morphological sequence is therefore not caused 
by disks but mostly by bulges.
It is also shown that elliptical galaxies and bulges of spiral
galaxies are unlikely to be in a single sequence.  We infer the
stellar mass density (in units of the critical mass density)
to be $\Omega=$0.0021 for spheroids, i.e.,
elliptical galaxies plus bulges of spiral galaxies, and
$\Omega=$0.00081 for disks.

\section{Introduction}

The galaxy consists of two distinct components, disks and bulges, and
how they formed is an outstanding problem in the galaxy formation.
The classical idea is that elliptical galaxies and bulges, which
are altogether called spheroids, formed when
infalling gas undergoes star formation during the initial collapse of
the system, and disks formed from dissipational collapse of the
rotating gas that is left-over after the initial free-fall
collapse (e.g., Eggen et al. 1962;
Sandage 1986). Until early 1970's it was widely taken that bulges
and elliptical galaxies belong to a single population and that
elliptical galaxies are spheroids that lack disks for some reasons.
This was because the shape and photometric properties of bulges
and ellpiticals are quite similar.

However, lines of kinematical evidence against the single population
hypothesis accumulated since late 1970's. Illingworth (1977) indicated 
that bright elliptical galaxies rotate slower than expected from
their ellipticities if their velocity
dispersion is isotropic. Kormendy and Illingworth (1982)
found that bulges of S0 and spiral galaxies rotate more rapidly
than bright elliptical galaxies, being consistent with rotationally
flattened oblate spheroids with isotropic velocity dispersion.
Davies et al. (1983) showed that faint elliptical galaxies also
rotate rapidly than bright elliptical galaxies and that no significant
difference is present between the kinematic properties of the bulges
and elliptical galaxies with comparable brightness.
Bender et al. (1988) discovered that the dichotomy of elliptical 
galaxies between slow rotators and rapid rotators is more clearly 
defined by the isophote shape than by luminosity; slow rotators 
always have boxy isophotes and often brighter than rapid rotators 
which always have disky isophotes.

Recently the more popular idea based on the hierarchical clustering 
scenario says that disk galaxies formed in the centre of dark matter 
halo as infalling gas collapses and spheroids are formed via 
violent mergers (Kauffmann et al. 1993; Baugh et al. 1996).
There is also a different view that bulges formed as a result of
secular evolution from the disk (Kormendy 1993; Athanassoula 2003;
Debattista et al. 2004; Kormendy \& Kenicutt 2004; Martinez-Valpuesta
et al. 2006; Fisher \& Drory 2008; Mendez-Abreu et al. 2008).
Lenticular galaxies in between elliptical 
and spiral galaxies have attracted much
attention as to their origins (e.g., Moran et al. 2007).
To study the problems of the formation
of spheroids and disks we must collect the statistics as to the
properties of the two components and study their regularities, in
particular how they vary across early to late types, for as many
galaxies as possible. The galaxy sample extracted from Sloan Digital
Sky Survey (York et al. 2000) based on CCD images provides us with a
good data base for this purpose at low redshift.

A number of methods have been used for the bulge-disk
decomposition. Traditional one-dimensional methods use a surface
brightness profile which is extracted from two-dimensional surface
brightness distribution by a variety of methods (e.g., Kormendy 1977;
Kent 1985; Simien \& de Vaucouleurs 1986; Kodaira et
al. 1987). Two-dimentional methods (e.g., de Jong 1996a; M\"ollenhoff
2004) are more popularly used recently and a number of semi-automatic
codes have been developed. They include GIM2D (Marleau \& Simard 1998;
Simard et al. 2002), GALFIT (Peng et al. 2002), BUDDA (de Souza et al. 
2004), GASPHOT (Pignatelli et al. 2006), and GASP2D (M\'endez-Abreu et
al. 2008).  Some of them were developed to tackle specific problems
for a specific sample of galaxies while others were intended to be
used in more general applications. They differ from each other in many
respects such as the number of components, fitting function for
respective components, minimization algorithm, and the degree of
automation.

In the profile decomposition of galaxies, we should keep in mind
a caveat that we do not know how well a single fitting function
represents the surface brightness distribution of the component
of real galaxies. Thorough investigation of the accuracy and
the robustness of these methods is yet to be made (e.g.,
Pignatelli et al. 2006). 
Existence of this many codes itself demonstrates the fact
that the most suitable method and code do depend upon
both the problem to be addressed and
quality and/or quantity of the data to be analysed.
Generally speaking, the two dimentional
method, while it should give more accurate decomposition, requires
accurate galaxy images with high signal to noise ratios, 
and is sensitive to the details of structures.
It does not always succesfully applies to a large scale sample
where relatively small size images are available.

In this paper we investigate systematic behaviours of
rudimentary properties, i.e., characteristic
scales, characteristic brightnesses and aboslute
magnitudes, of bulge and disk components along the Hubble 
sequence by means of a one-dimesional method based not on the
surface brightness profile but on the growth curve
of galaxies using a large homogeneous sample.
We admit that our decomposition may not be quite accurate
galaxy by galaxy basis, but we believe it gives useful
data and provides us with the information for galaxy science
so far missing that should be associated with the SDSS data.

\section{Galaxy sample}

We take a sample of 1600 galaxies with r$<$15.9 and measured 
redshift taken from the equatorial stripe
$145.15^\circ<\alpha<235.97^\circ$, $|\delta|<1.26^\circ$
in the northern sky (229.7 sq. deg.),
for which morphological classification was
carried out by visual inspections. This is an earlier version of
the sample in the catalogue given by
Fukugita et al. (2007), which contains 2253 galaxies with  $r\leq16$
in the same region (1866 are given redshift).

Among the 1600 galaxies, 1044 are classified as S0 to Sc.
From the 1044 we discard 239 galaxies too
close to edge-on (b/a$<$0.3), 14 galaxies that have bright stars
or galaxies overlapped with the galaxy images, 15 galaxies
with a lack of growth curve data for some outer parts, 3 galaxies 
with a lack of confident redshift.
We also dropt 20 galaxies that are not suitable for
accurate photometry, either located close to
the edge of the survey area or contain saturated pixels.
The remaining 753 galaxies were subject to the bulge-disk
decomposition and 737 yielded satisfactory results.
The analysis in this paper is thus based on 737 S0-Sc galaxies.
The sample we use here does not show any particular
bias compared with the $r<16$ morphologically-classified sample
of Fukugita et al. (2007) except that there are
some missing galaxies close to the faint end in our sample.

We use $r$-band images drawn from Data Release
3 of SDSS (Abazajian et al. 2005),
which do not differ from images in the later data
releases for galaxies that concern us.
Galactic extinction were corrected according to
Schlegel et al. (1998).
The number of galaxies for each morphological type used in our 
analysis is presented in Table 1 together with other statistics 
that will be discussed in this paper.

We refer the reader to the other publications for
descriptions of the SDSS related to our study: Gunn et al (2006) for
the telescope, Gunn et al. (1998) for the photometric camera, Fukugita
et al. (1996) for the photometric system, Hogg et al. (2001) and Smith
et al. (2002) for external photometric calibrations, Pier et al.
(2003) for astrometric calibrations, and Strauss et al. (2002)
for spectroscopic target selection for galaxies.  We also refer to 
Abazajian et al. (2003; 2004) and Adelman-McCarthy et al.
(2006; 2007; 2008) for other data
releases from the SDSS, which discuss the successive improvement of
the pipelines used to derive the basic catalogues.
We use $h_{70} = H_{0}/(70$ km s$^{-1}$ Mpc$^{-1}$)
unless mentioned.

\section{Growth curve fitting and the bulge-disk decomposition}

The growth curve is the flux integrated within
a circular aperture in units of magnitudes
as a function of the aperture. It has traditionally been used to
estimate total magnitudes (e.g., de Vaucouleus et al.; RC2).
The growth curve should in principle contain the information on
profiles of the bulge and the disk.
The method used here is not new and the code was developed
and tested in Okamura et al. (1999) using simulations and
a sample of real galaxies available at that time.
Several essential points relevant to the present study
are summarized below.

We assume that galaxies are represented by two components,
bulges and disks, and their surface brightness $I$ is described 
with the de Vaucouleurs law
\begin{equation}
{\rm log} (I/I_{e,B}) = -3.33 \left[ (r/r_{e,B})^{1/4}  - 1 \right]
\label{eq:deV}
\end{equation}
for bulges and the exponential profile
\begin{equation}
{\rm log} (I/I_{e,D}) = -0.729 \left[ (r/r_{e,D})  - 1 \right]
\label{eq:exp}
\end{equation}
for disks, where for the two respective components
$r_{e,B}$ and $r_{e,D}$ are the effective radii
within which half the total flux of each
component is contained, and $I_{e,B}$ and $I_{e,D}$ are surface
brightness at these effective radii. 

Departures from the de Vaucouleurs profile 
(\ref{eq:deV}) are often argued, in particular for
late type spiral bulges (van Houten 1961; Andredakis et al. 1995;
Courteau et al. 1996; Trujillo et al. 2001;
M\"ollenhoff 2004; Aguerri et al. 2004, 2005; Mendez-Abreu et al. 2008),
that the profile of the bulges are somewhat
steeper than the de Vaucouleus profile and an arbitrary power of $r$
is introduced in the exponent for a general profile (S\'ercic 1968).
Fittings with a general S\'ercic profile, however, is not stable
unless the image has high signal-to-noise ratio over 
sufficiently wide dynamic range,
causing the degeneracy among parameters, especially between the scale
length and the power index (see, e.g., Trujillo et al. 2001).
Our study is based on the sample of images of relatively low
(at least in terms of conventional bulge-disk decomposition)
signal-to-noise ratio with a limited length scale, and it is 
hard to discern S\'ercic-type
powers in the bulge component. Hence, we avoid introducing an
extra parameter for the bulge profile that controls the power of $r$.
Parameters derived from the fitting are different for different 
fitting functions used but they represent virtually the identical
physical property if the function is not too far from
the reality (see Appendix of Kormendy 1977; Fig.1 of Graham 2001a).
The sytematic behavior of the parameters along
the Hubble sequence, which we focus on in this study, depends only weakly
on the specific choice of a particular fitting function.

We note that a two-dimensional fitting for the bulge with the S\'ercic
profile shows that 2/3 of galaxies have $n=4$, i.e., the de Vaucouleurs
profile as the best fit solution (Tasca \& White 2005), so that the 
error for the global mean arising from the assumption enforcing 
$n=4$ is not too large.

Another issue of concern is the effect of bars. Our growth
curve method is a one-dimensional method based on a series of
{\it circular} apertures. Accordingly, information on the elliptical
surface brightness distribution of a bar is 'degenerated' onto
an equivalent circular surface brightness distribution of a
'hypothetical' bulge. This means that in case of a barred galaxy
we regard the bar plus bulge as a single entity, which we call
bulge here. This point is discussed further in section 4.

For the growth curve method to work properly, it is essential to
include the effect of finite seeing, which we take
to be double Gaussian that is the SDSS default
and parametrise for our purpose its full width half maximum $w_s$ 
as its ratio to the effective radius of the disk
\begin{equation}
\zeta=\log(r_{e,D}/w_s)\ .
\label{eq:seeing}
\end{equation}
The seeing parameters are catalogued for each galaxy in the SDSS
data base.

For our application we prepare templets for 11 grid points
for each of the three parameters, $B/T=0-1$,
$\eta=\log(r_{e,B}/r_{e,D}), -1.2<\eta<0.8$ and
$0<\zeta<2.0$. The ranges of the parameters are chosen so that
the results are well covered in these ranges. We then compute
\begin{equation}
\chi^2=\sum_{i=3}^{N} w_i \Big[ m(\log r_i)-m_{\rm tot}-
\Delta m\left(\log\frac {r_i}{r_e^T}\right)\Big]^2
+\frac{1}{4}\Big[\log r_e-\log r_e^T\Big]^2,
\end{equation}
where the templet was swept over all 1331 grids with the parameters
$B/T$, $\eta$, $\zeta$, together with two free parameters, $m_{\rm
tot}$ and $r_e$ searched to give the best fit to the growth curve data
for each templet. Here, $m_{\rm tot}$ is the total magnitude and $r_e$
is the half light radius of the galaxy, while $r_e^T$ is the half
light radius of the assumed templet.  The weight factor is taken to be
$w_i=\log r_i$.  The last two terms are added to avoid a fake fit with
a rather unrealistic templet, to ensure that the best fit parameter
for the effective radius is close to the one adopted as a templet. We
use the data at 13 apertures from 1.03 to 263 arcsec with
approximately a geometric sequence by a factor of 1.5, measured by the
photometric pipeline of SDSS (PHOTO in short, see Stoughton et
al. 2002).  We have removed the two innermost growth curve data points
from PHOTO, those at 0.22$''$ and 0.68$''$, from the fitting, as they
are substantially smaller than median seeing, 1.3$''$, and strongly
susceptible to seeing.
$w_s$ is known
for each galaxy, so that $r_{e,D}$ and $r_{e,B}$ are obtained from
$\zeta$ and $\eta$ of the templet that gives the best fit to growth
curves.
We apply the K correction to the image, though all galaxies have
small redshift $z < 0.1$, taken from Fukugita et al. (1995)
assuming bulges to have elliptical color and disks Scd color.
Their K corrections are consistent with the mean of those calculated
using individual spectra (Blanton et al. 2003), and anyway the 
redshifts are so low that errors are negligible.

Extensive simulations show that this growth curve method can be
used to determine bulge and disk parameters and bulge-to disk
luminosity ratios provided that the point spread function is
accurately known and signal-to-noise ratio is modest, say S/N$\ge30$
(Okamura et al. 1999).  It is shown that the accuracy of
the derived parameters depends upon inclination,
bulge-to-disk ratio, surface brightness, available image area, and
also other factors, in addition to S/N.  
It is, therefore, in general difficult to quote a few
numbers to represent the robustness of the method. 
Another example of similar presentation can be seen in Pignatelli et
al (2006), where the robustness is estimated as a function of the
threshold area, instead of S/N, of galaxy images.  
The present sample is limited to galaxies with $r<15.9$, which
roughly corresponds to S/N$>200$ (see Figure 2 of Okamura et al. 1999), 
and does not include highly inclined (b/a$<$0.3) galaxies.
Accordingly, we can expect a reasonable robustness for the 
present sample. Fit to each of the sample galaxies is visually 
examined.

For a verification of the fit, we show
the difference between the Petrosian magnitude
measured by the SDSS photometric pipeline and
the total magnitude obtained from the present fit as a
function of $B/T$ in Figure 1 (a), and
the effective radius $r_e$ with respect to the Petrosian
half flux radius $r_{P50}$ as a function of
$B/T$ in Figure 1 (b), where the plotted radii
from the fit are corrected for finite seeing
to compare with the measured Petrosian radii.
Note that we expect $m_{P}-m_{\rm tot}$=0.221
and 0.007 for the ideal de Vaucouleus and
exponential profiles that are located face on,
which are indicated by horizontal lines in the figure.
For inclined galaxies, these offsets are slightly ($\sim 5\%$)
smaller. The data for magnitude offsets in Figure 1 (a) are
located mostly in between 0 and 0.2 mag, with a trend
for an increase to 0.2 mag towards a larger $B/T$ in agreement
with our expectation.

For the effective radii, we expect $r_{P50}/r_{e}=0.713$ for de
Vaucouleus galaxies and $r_{P50}/r_{e}=0.993$ for exponential galaxies
for the face-on case.  Our data in Figure 1 (b) that are located
mostly between 1 and 1.3 with relatively larger values for bulge
dominated galaxies as expected.  There are a small number of cases
where deviations are significant. We have examined images of 
those cases and found that they happen occasionally for
large-size late-type galaxies, for which PHOTO gives too small radii
which do not seem to be correct.  This is probably due to
errors in the measurement by PHOTO which are caused by
too marked contrasts with bright bulges.  Those data points in Figure
1 (a), whose Petrosian magnitudes are somewhat too dim,
also correspond to the deviants in (b) and the same reason is suspected.
Together with the figure for total magnitudes our result means that
the fit gives a reasonable value for both total flux and effective
radius, and our assumption for galaxies that are represented by
exponential plus de Vaucouleurs profiles works reasonably well.

We adopt the morphological type index $T=0$ for E,
$T=1$ for S0, $T=2$ for Sa, $T=3$ for Sb, $T=4$ for Sc and $T=5$ 
for Sd, since the T index as detailed as that in 
the RC2 is not warranted for both our catalogue
and purpose here. Galaxies of classes with half integer $T$ are 
grouped into the neighbouring later class
except for $T=0.5$ (E/S0), which shows unclear sign of
disks and is discarded in our decomposition analysis.
The morphological index given in Fukugita et
al. (2007) is based on visual inspections of $g$ band images
by several independent classifiers in
reference to prototypes presented in {\it Hubble Atlas of Galaxies}
(Sandage 1961). The mean index  by the several classifiers is 
given in the catalogue, which we take in the present work.

Our prime interest is to study the change of properties of bulge and
disk against morphology, but our sample is large enough to attempt to
study the change of properties of the bulge and the disk against
luminosity. We make three subsamples $-23<M_r<-22$, $-22<M_r<-21$ and
$-21<M_r<-20$, dividing the sample into three luminosity groups; they
stand for nearly luminosity limited samples with varying distance
limits.

\section{Results}

\subsection{Bulge to Total Luminosity Ratio}

Figure 2 presents the bulge to total luminosity ratios as a function
of morphological type index $T$.  The mean and dispersion are shown by
error bars for each morphological type, where $T$ runs from 1 for S0
(excluding E/S0)  to 4 for Sc (including Sbc).  
In order to assess the effect of bars discussed in 
section 3, we examine all the galaxy images and
classified them into barred and non-barred. 
Except for T=1 (S0), where only three barred galaxis are
present, no systematic difference is found between barred 
and non-barred galaxies.
This is probably due to the fact that the elliptical
structure of bars disappears in the growth curve obtained
in circular apertures as mentioned in section 3.
Laurikainen et al. (2007) found some difference in 
$B/T$ versus Hubble type between early-type barred and 
non-barred spiral galaxies. However, their data are based
on near infrared images and a direct comparison is 
inappropriate.

One can see a good
correlation showing that an earlier type shows a larger $B/T$, 0.64
for S0, decreasing for late types to 0.19 for Sc.  This agrees
broadly with the results given by a number of authors, e.g.,
Kent (1985), Simien and de Vaucouleurs (1986), Kodaira et al. (1986),
and others, although an accurate comparison needs a translation as to
different color bands used by respective authors. 
The numbers for S0 and Sc, 0.64 and 0.19, are,
for instance, compared with $0.75{+0.1\atop -0.3}$ and
$0.07{+0.15\atop -0.05}$ of Kent (1985), who used Thuan-Gunn $r$ band
that is close to ours.  The dispersion for $B/T$ in each class,
approximately $\pm 0.2$, denoted by error bars, however, is larger
than the interclass differences. This also agrees with what is known
from analyses made in the past: the scatter in each class is larger
than the difference of the average among different classes.  This
means that, while $B/T$ is well correlated with morphological types,
we cannot replace the morphological types for individual galaxies
with $B/T$, although $B/T$ provides a convenient measure and
is often used to classify morphological types (e.g., Tasca \& White
2005) especially in theoretical modelling (e.g., Baugh et al. 1996).
It is tempting to ask if this systematic variation of the bulge to
total luminosity ratio is ascribed to the property of the bulge or the
disk, or both. This question is answered in the subsections that follow.

\subsection{Properties of Bulges}

The two key parameters that characterise the bulge are the effective
radius $r_{e,B}$ and $\mu_{e,B}$ which is surface brightness at
$r_e$.  It was shown that these parameters for elliptical
galaxies obey some $\mu_e-r_e$ relation (Kormendy 1977).
We show in Figure 3 $\mu_{e,B}$ and $r_{e,B}$
for the bulge component with different symbols meaning
different morphological types.
Dotted lines indicate bulge luminosities being constant,
$-M_B=16, 18, 20$, and 22.
The solid line shows the
$\mu_e-r_e$ relation, or Kormendy's relation, for early-type
galaxies derived from the SDSS galaxy sample,
$\langle\mu\rangle_e=2.04\log r_e+18.7$ (Bernardi et al. 
2003).\footnote{
The original $\mu_e-r_e$ relation given in Kormendy (1977)
was based on $\mu_e$, the surface brightness {\it at} $r_e$.
Some later studies, however, use $\langle\mu\rangle_e$, the
mean surface brightness {\it within} $r_e$, which is by definition,
brighter than $\mu_e$. When comapring the Kormendy relations
based on the different definition of the surface brightness
parameter, care must be taken of the offset; for the
$r^{1/4}$ profile, $\mu_e - \langle\mu\rangle_e$ = 1.39.
The line drawn in Fig. 3 is $\mu_e=2.04\log r_e+20.09$.}
The region where most of 9000 \lq early-type
galaxies' of SDSS sample\footnote{
The selection of early-type galaxies by Bernardi et al. (2003) 
is rather rudimentary to deal with a large sample.
It contains not only E and S0 galaxies but also many Sa galaxies, 
when compared it with visually classified sample. The sample, 
however, is certainly rich in E and S0 galaxies, and we 
expect that statistical quantities derived from the sample give
a reasonable approximation. We also note that the sample suffers
from a significant incompleteness for early-type 
galaxies fainter than $M_r\simeq -21$.}
is distributed is shown by the dashed ellipse.
The derived slope looks shallower than that for the
original Kormendy relation, $\mu_e=3.02\log r_e+19.74$
($B$ band). The slope of the Kormendy relation
varies with brightness of galaxies, steeper for
fainter galaxies (Nigoche-Netro et al. 2008). The sample
of Kormendy (1977) spans over M$_B\sim-20$ to $-22$, while
the SDSS sample extends from 
M$_R\sim-19$ to $-24$.

It is clear that $(\mu_e, r_e)$ of bulges are distributed in
a region much wider than the \lq SDSS early-type galaxies' and 
as a whole they do not follow the relation for SDSS early-type 
galaxies. We note, however, that ($\mu_e, r_e$) of bulges of S0 
galaxies closely follow the relation for the SDSS early-type 
galaxies in the overlapped region (dotted ellipse). This is 
reasonable because the SDSS sample includes S0 galaxies
(Bernardi et al. 2003).

Bulges of later-type spiral galaxies
$(\mu_e, r_e)$ are distributed along lines that are significantly
steeper, nearly along the line of constant luminosity,
$I_e\,r_e^2$=constant, or
$\mu_e=5\log r_e+$constant, with a significantly scatter larger
than S0 bulges.
It is known that there is a dichotomy in elliptical
galaxies in terms of kinematical structure: slow rotators
versus rapid rotators (e.g., Kormendy and Illingworth 1982;
Davies et al. 1983; Davies and Illingworth 1983).
Slow rotators have boxy isophotes
while rapid rotators have disky isophote (Bender et al. 1988).
This dichotomy is also closely related with luminosty;
bright ellipticals are often slow rotators while faint ellipticals
tend to be rapid rotators with the boundary at $M_B\sim-20.5$ mag.
The $\mu_e-r_e$ relation of rapid rotators is not well known
since even the SDSS sample, by far the largest sample of early-type
galaxies, contains only a small fraction of ellipticals
fainter than  $M_B\sim-20.5$ ($M_R\sim-22$). 
Whether or not the bulges of late-type spiral galaxies
and rapidly rotating faint ellipticals
follow a similar $\mu_e-r_e$ relation is at present
an open question.

If we consider the mean $(\mu_e, r_e)$, we see
a trend (see Figure 4) for
different morphological classes: bulges for late type
spirals have surface brightness
dimmer than that for early types, whereas the effective
scale length of bulges differ little.
The bulge surface brightness of Sc spirals is dimmer by 2 mag
arcsec$^{-2}$ than that for S0 or Sa. Surface brightness of
Sa bulges, however, is nearly the same as that for S0 bulges.
The bulge luminosity for late-type spirals are also lower by 2 mag
than that for S0's.

It is noted in passing that no systematic difference is found
between barred and non-barred galaxies. When the plot is
made separately, they are distributed
over the same area in the $(\mu_e, r_e)$ plane.

\subsection{Properties of Disks}

A similar figure for ($\mu_e, r_e$) is shown for disks in Figure
5. The data are apparently distributed more clustered in a narrower
region than for bulges: a rough trend $I_e\,r_e^2\sim$ constant
is still visible, but the data are distributed in narrower
ranges. The mean surface brightness is $\mu_{e,D}=22.06$ mag
arcsec$^{-2}$ with the dispersion
0.96 mag arcsec$^{-2}$.
No systematic difference is found here either between 
barred and non-barred galaxies in their distribution in
the ($\mu_e, r_e$) plane.

Figure 6 shows the mean and dispersion of ($\mu_e, r_e$) for each
morphological type, indicating that the properties of disks change
little against morphologies; at least the change is not systematic
along the morphological sequence. The disk luminosity also differs
little across S0 to Sc. For example, properties of disks for S0 and Sc
galaxies differ very little: the difference among different morphology
classes is much smaller than the scatter from galaxy to galaxy in one
class. Surface brightness of S0 may be slightly fainter than that of
later spiral galaxies but at most only be 0.5 mag arcsec$^{-2}$.
The mean surface
brightness $\langle \mu_e\rangle$ is located from 21.5 to 22.5 mag
arcsec$^{-2}$ with no systematic trend visible against the
morphological sequence. This fact may also be interpreted as
indicating the universality of disks in general, and reinforces the
universality found by Freeman (1970) based on a handful of spiral
galaxies available in those days (see also de Jong 1996b).

It is interesting to note that the morphological type dependence, as
we have seen with bulges, almost disappears for disks.  Disks in both
early- and late-type spiral galaxies, including S0's, are similar with
nearly constant surface brightness and scale length independent of
the disk galaxy morphology.

\subsection{Summary: bulge and disk properties vs. morphological type}

The most conspicuous fact is that the bulge to disk or bulge to total
luminosity ratio depends on morphology of spiral galaxies. Later-type
spiral galaxies are more disk dominated in agreement with widely
accepted concept. We have found that the properties of disks,
including luminosity, do vary little with morphology. This implies that the
property that varies with morphology is the bulge; in fact bulge
luminosity is the main variable that controls
morphology. A large and conspicuous bulge means the galaxy being an
early type.  In contrast, the properties of disks are nearly
universal and depend little on morphology.

These systematic behaviors of bulge and/or
disk parameters were found in previous studies
based on various samples of galaxies (e.g.,
Kodaira et al. 1986; de Jong 1996a; Graham 2001a,b;
M\"ollenhoff \& Heidt 2001; Trujillo et al. 2002; Aguerri et al. 2004;
Laurikainen et al. 2007; M\'endez-Abreu et al. 2008).
The size of these samples ranges several tens to 
about two hundreds and some are limited to cluster members.
Our study confirms the behaviors that have been referred to in
the literature based on a much larger homogeneous magnitude-limited
sample in the field. An accurate quantitative comparison needs
translation as to different color bands as well as
specic bias arising from different
methods adopted by different authors.

The systematic behaviors found here and in some previous
studies may imply that the formation of disks takes place independent of
bulges. The infall of intergalactic gas, for example, takes place
irrespective of bulge properties.
Some self-regulating mechanisms are suspected to be at
work in disk that limit the accumulation of too many stars per
area of the disk.

We do not see any trend that later type disks have higher surface
brightness. In particular surface brightness of S0 disks
differs little from that of Sa disks,
in contrast to the claim that Sa disks have surface brightness
much brighter than S0 disks (Sandage 1986), as expected in the
monolithic collapse scenario.  Surface brightness of Sa and that of Sc
are also nearly identical on average.  We see no signatures that Sb-Sc
disks are less luminous than that of S0-Sa, which might be resulted if
disk stars are transported to bulges by secular formation of bulges
from disks in late type spiral systems (Kormendy 1993; Kormendy \&
Kennicutt 2004).

It is also interesting to note that surface brightness of S0 galactic
disks differ little from that of other spiral galaxies.  The majority
of S0's disks are unlikely to be a result of stripping of spiral
galaxies (Larson et al. 1980) or faded spiral galaxies (Bedregal et
al. 2006): S0's, at least for their majority, are not a result from
spiral galaxies that lost substantial disk ingredients.

\subsection{Correlation between scale lengths}

The evidence has been discussed that the disk scale length correlates
with the bulge scale length and it is taken as evidence that suggests
disk origin of bulges by secular evolution of disks
(Courteau et al. 1996; M\'endez-Abreu et al. 2008):
galaxies with large size bulge may necessarily have
large size disks. In our sample we have not seen any
particular correaltion between the two scale lengths. The
correlation, if any, is week.

The disk scale lengths are distributed dominantly between 2 to 10 kpc, 
and the bulge scale lengths between 0.2 to 6 kpc.
The ratio  $r_{e,B}/r_{e,D}$ is
distributed in the range of $1-10$ with very broad peaks whose
center is around 0.3  (see Figure 7 for
histograms for $r_{e,B}/r_{e,D}$). This might be taken as the evidence
of correlations between the two scale lengths
but in fact does not mean the presence of a particular correlation
between the two scales: it is a result of a fact that the two scales
are distributed independently in narrow ranges.
We do not either observe close correlations between quantities
characterizing bulges and disks, implying that disks formed
independently of the details of bulges.

\subsection{Properties of disks and bulges as luminosity varies}

We carried out the same analysis with the three luminosity group 
samples. The trends we observed in this section for the total sample
are still visible with the luminosity grouped samples and the overall
trends differ little from that we have seen for the total sample.
Three panels in Figure 8 give the change against total luminosity 
of galaxies.

First we see (Figure 8 (a)) that the change of the bulge-to-total flux
ratio against the total luminosity. The change is only little, $\sim
20$\%, even for S0 galaxies, where the bulge largely controls the
total luminosity. The change is hardly discernible for later type
spiral galaxies.
This result means that luminosities of both bulges and disks change as
total luminosity changes in a similar way (and in a lesser degree for
disks), so that the ratios stay nearly at constants.

Bulges of fainter galaxies are also fainter by the same amount, but
the reason is not uniform (see Figure 8 (b)): in early type disk
galaxies, S0 and Sa, this change arises more from the decrease of the
size, fainter galaxies having smaller bulges, but in later types
dimmer bulge magnitude is ascribed primarily to dimmer surface
brightness rather than smaller bulge sizes, which remain a constant
independent of luminosity.

Luminosities of disks change as the total luminosity changes in nearly
the same way (less in earlier systems). The change of disk luminosity
is mostly caused by the decrease of the size (effective radius) while
surface brightness at the effective radius stays nearly unchanged: the
change of $\mu_e$ is less than $-$0.3 mag arcsec$^{-2}$ for a 2 mag
change of disk luminosity (the change of surface brightness takes
place in the opposite direction). See Figure 8 (c).

\section{Luminosity density}

Nakamura et al. (2003) estimated that the luminosity densities
of early and spiral-type galaxies galaxies to be
\begin{eqnarray}
{\cal L}_r({\rm E+S0})=0.43 \times 10^8 h_{70} L_{\odot}\
                      {\rm Mpc^{-3}},\\
{\cal L}_r({\rm S})=0.96 \times 10^8 h_{70} L_{\odot}\
                      {\rm Mpc^{-3}}.
\end{eqnarray}
These values and all the numbers upto and including eq. (10) 
should be multiplied by 1.29 for the global values to
correct for the underdensity of galaxies in the northern equatorial
stripes where the luminosity function (and also present work) was
derived from. Assuming that the shape of the
luminosity functions change little within rough classes of morphology
according to Nakamura et al.  (2003) and using their luminosity
densities up to the normalisations, we infer that
\begin{eqnarray}
{\cal L}_r({\rm S0,bulge})=0.18 \times 10^8 h_{70}\ L_{\odot}\
                          {\rm Mpc^{-3}},\\
{\cal L}_r({\rm S0,disk})=0.10 \times 10^8 h_{70}\ L_{\odot}\
                          {\rm Mpc^{-3}},
\end{eqnarray}
where E : S0+E/S0 = 0.36 : 0.64 from Fukugita et al. (2007), 
and B/T of S0 is 0.64 from Fig. \ref{fig2}, and the
luminosity fractions given in section 3 are assumed to be independent
of luminosity and are used to compute the bulge
and disk contributions. The luminosity density of elliptical 
galaxies is
${\cal L}_r({\rm E})=0.15 \times 10^8 h_{70}\ L_{\odot}\ 
{\rm Mpc}^{-3}$.

For spiral galaxies Nakamura et al. give
${\cal L}_r({\rm S})= 0.70 \times 10^8 h_{70}\ L_{\odot}\ {\rm Mpc^{-3}}$
for S0/a to Sb galaxies and
$0.26 \times 10^8 h_{70}\ L_{\odot}\ {\rm Mpc^{-3}}$
for Sbc to Sd galaxies.
Assuming the bulge to total luminosity ratios for Sd-Sdm (which
contribute by only 13\% to the luminosity density of Sbc-Sdm) being
equal to that for Sc galaxies, we use 0.19 for B/T of Sbc to Sd galaxies.
We also assume B/T of S0/a to Sb galaxies to be 0.43 as an
approximate mean in the morphology range.
These values enable us to infer the luminosity densities
for spiral galaxies in the same way and we obtain
\begin{eqnarray}
{\cal L}_r({\rm S+S0,bulge})=0.53 \times 10^8 h_{70}\ L_{\odot}\ 
                             {\rm Mpc^{-3}},\\
{\cal L}_r({\rm S+S0,disk})= 0.71 \times 10^8h_{70}\ L_{\odot}\ 
                             {\rm Mpc^{-3}}.
\end{eqnarray}
Therefore contributions to the luminosity density from
disk, bulge and elliptical galaxies are
0.51:0.38:0.11, respectively.

This result may be compared with 0.54: 0.14: 0.32 from Tasca and White
(2004). The disagreement in the spheroidal contributions is ascribed
to the fact that a significant amount of S0 galaxies are counted as
pure bulge systems, i.e., elliptical galaxies, in Tasca and White.
Intrinsically bright galaxies are very often regarded as pure
bulge systems with nearly edge-on disks greatly under-represented in the
application of Gim2D code.
When spheroids include elliptical galaxies and bulges, the relative
contributions to disk and spheroids 0.53:0.47 in our analysis agrees
with the fraction given by Tasca and White.

If we use a result of stellar population 
synthesis (Bruzual \& Charlot
2003) assuming two populations in the Universe --- spheroids and disks
--- constrained with the mean colors of SDSS galaxies (Nagamine et
al. 2006), we have the average stellar mass to light ratios,
$\langle M_*/L_r\rangle=3.2$ for spheroids and $1.2$ for disks,
or correspondingly
$\langle M_*/L_B\rangle=5.6$ and $1.2$, respectively,
if the more familiar $B$ band is adopted.
Using these mass to light ratios and including the fraction
of 1.29 correction and $h_{70}=1$
we estimate the stellar mass density
\begin{eqnarray}
\Omega_{\rm spheroids, star}=0.00207,\\
\Omega_{\rm disk, star}=0.00081.
\end{eqnarray}
i.e., the total stellar mass density $\Omega_{\rm star}=0.0029$
and the ratio of two mass densities
$\Omega_{\rm spheroid, star}/\Omega_{\rm disk, star}=2.6$.
The latter is somewhat smaller
than the ratio in Fukugita et al. (1998), who gave 3, but
substantially larger than 0.77 by Benson et al. (2002)
and 0.75 by Driver et al. (2007). 

\section{Conclusions}

We have carried out bulge-disk decomposition for a modestly large
sample of galaxies derived from a morphologically classified sample of
the Sloan Digital Sky Survey with the use of groth curve fitting.  We
demonstrated that growth curve fitting method
works as expected and studied properties of
bulges and disks thus decomposed as functions of morphology and
luminosity. We
endorsed the well-known trend that the importance of bulge decreases
systematically from early to late types, but we have shown that this
is dominantly due to the variation of bulges as the morphological type
changes.  In contrast, we have shown that the properties of disks are
nearly universal, and depend only weakly on morphology
classes.  We do not see any systematic trend 
of disks that changes with morphology of disk
galaxies. In spite of a good correlation between the bulge to total
luminosity ratio and the morphological type, the galaxy to galaxy
scatter of the former is so large that one cannot replace the
conventional morphology type with the luminosity ratio.

While we have a number of different scenarios for disk and bulge
formation, definitive predictions that can be compared with our
analysis are not readily available. However a number of predictions or
likely results seem to be disfavoured. For example, the monolithic
scenario that bulge formed from infall of gas and disk formed
dissipational collapse of gas left-over the initial collapse favours
brighter disks for later type morphologies. This is not supported by
our data.  The model that S0 forms as a result of stripping of spiral
disks or spiral galaxies of faded disks are not favoured either, since
disk properties of spirals galaxies are all similar including S0's. We
do not observe any conspicuous change in disk properties between early
and late type spiral galaxies, which would be expected if late type
spiral bulges are a product of secular evolution from disks while
earlier types are from major mergers.  Furthermore we do not see
correlation between bulge and disk properties.  The universal disks
that we have seen may be more consistent with the naive idea that
disks are later additions by accretion of intergalactic material
falling onto bulges where accretion took place independ of bulge
properties but with some self-regulating mechanism that limits the
column density of stars in the disk at work for disks.

We also noted that the properties $(\mu_e,r_e)$ of bulges of spiral
galaxies and bright elliptical galaxies are different
as obeying different relations, which implies that bulges and
bright elliptical galaxies are
unlikely to be on a single sequence.

\acknowledgments

We thank Masayuki Tanaka for his help in the application of
the rather old growth curve fitting code to the present SDSS sample,
and Ivan Baldry for comments on the early version of the draft.
This work was supported in Japan by Grant-in-Aid of the Ministry of
Education. MF received support from the Monell Foundation at the
Institute for Advanced Study.

Funding for the SDSS and SDSS-II has been provided by the Alfred P. 
Sloan
Foundation, the Participating Institutions, the National Science 
Foundation,
the U.S. Department of Energy, the National Aeronautics and Space
Administration, the Japanese Monbukagakusho, the Max Planck Society, and
the Higher Education Funding Council for England. The SDSS Web Site is
http://www.sdss.org/.
The SDSS is managed by the Astrophysical Research Consortium for the
Participating Institutions. The Participating Institutions are the
American Museum of Natural History, Astrophysical Institute Potsdam,
University of Basel, Cambridge University, Case Western Reserve 
University,
University of Chicago, Drexel University, Fermilab, the Institute for
Advanced Study, the Japan Participation Group, Johns Hopkins University,
the Joint Institute for Nuclear Astrophysics, the Kavli Institute for
Particle Astrophysics and Cosmology, the Korean Scientist Group, the
Chinese Academy of Sciences (LAMOST), Los Alamos National Laboratory,
the Max-Planck-Institute for Astronomy (MPIA), the Max-Planck-Institute
for Astrophysics (MPA), New Mexico State University, Ohio State 
University,
University of Pittsburgh, University of Portsmouth, Princeton 
University,
the United States Naval Observatory, and the University of Washington.

\vspace{2cm}

\noindent
{\bf Appendix A: Stellar population synthesis}

The bulge disk decomposition enables us to infer color of
galaxies with two population of stars, disk stars and bulge stars. We
approximate them as delayed exponential model, i.e., star formation
being given by $\dot\rho_*=A(t/\tau)\exp(-t/\tau)$ (Searle et al 1973;
Nagamine et al. 2006). We may adopt the parameters given by Nagamine
et al.  (2006) which gives global galaxy colors as our fiducial
choice without a further adjustment: $\tau_d=4.5\,{\rm Gyr}$ for
disk and $\tau_b=1.5\,{\rm Gyr}$ for bulge.  
Using our bulge to disk ratio, we calculate
$g-r$ and $u-g$ colors of galaxies with various morphological types
as shown in Figure 9. Here we take the stellar population synthesis of
Bruzual and Charlot (2003). The figures also give the mean and
variance estimated in the morphologically classified galaxy sample
(Fukugita et al. 2007), showing reasonable agreement between the two.

This is an example of the significant use of the bulge disk
decomposition.  These figures mean that one can construct a
qualitative model of galaxies consistent with the observation using
the simple bulge disk decomposition given in this paper.  With this
model we can also calculate stellar mass to light ratio as
\begin{equation}
M_*/L_r=1.18+2.01(B/T),
\end{equation}
where $B/T$ is the bulge fraction of luminosity given in Table 1,
and Chabrier's initial mass function (Chabrier 2003)
is used to calculate the stellar mass.


\begin{deluxetable}{lcccccccc}
\tablecolumns{7}
\tablewidth{0pc}
\tablecaption{Properties of galaxies for our sample}
\tablehead{
\colhead{Hubble type} & \colhead{S0} & \colhead{S0/a-Sa} 
&\colhead{Sab-Sb}&\colhead{Sbc-Sc}}
\startdata
$T$(ours)&1&2&3&4\cr
nrs. of galaxies & 183 & 158 & 184 & 212 \cr
$B/T$ & $0.64\pm0.19$ & $0.49\pm0.24$& $0.29\pm0.20$ & $0.19\pm0.14$ \cr
$r_{e,B}$ ($h_{70}^{-1}$ kpc)&$2.95\pm2.30$ & $2.56\pm2.81$
&$2.76\pm2.56$ &$3.02\pm2.82$ \cr
$\mu_{e,B}$ (mag arcsec$^{-2}$)& $20.31\pm1.26$ &$20.62\pm1.53$
&$21.53\pm1.97$  &$22.15\pm1.91$ \cr
$r_{e,D}$ ($h_{70}^{-1}$ kpc)&$7.20\pm4.56$ &$5.51\pm3.38$
&$5.25\pm2.98$ &$6.71\pm3.15$ \cr
$\mu_{e,D}$ (mag arcsec$^{-2}$)&$22.56\pm1.14$ &$22.04\pm1.01$ &
$21.67\pm0.71$ &$21.96\pm0.65$ \cr
$\log[r_{e,B}/r_{e,D}]$&$-0.45\pm0.33$ &$-0.44\pm0.42$ &$-0.39\pm0.48$
&$-0.45\pm0.43$\cr
\enddata
\label{table:counts}
\end{deluxetable}

\newpage
\begin{figure}
\plotone{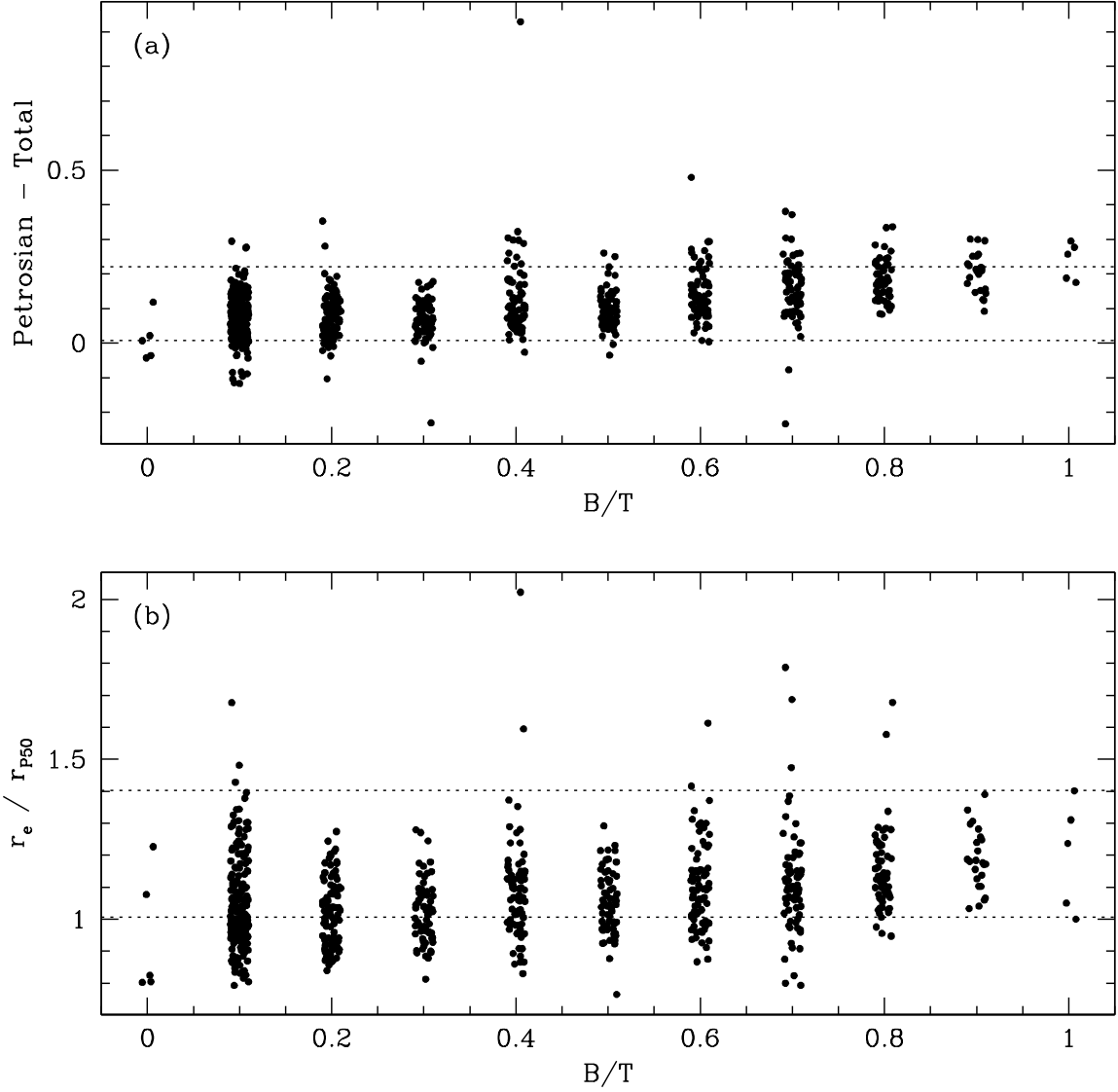}
\caption{(a) Offsets between $m_{tot}$ obtained from our fitting
and $m_{P}$ the Petrosian magnitude estimated from the photometric
pipeline of the SDSS, and (b) offsets between the effective radius
$r_e$ from our fit and the SDSS output of 50\% Petrosian radius
$r_{P50}$, both as a function of the bulge to total luminosity ratio.
The dashed lines are offsets expected for galaxies with the pure
de Vaucouleurs and exponential profiles located face-on.
\label{fig1}}
\end{figure}

\begin{figure}
\plotone{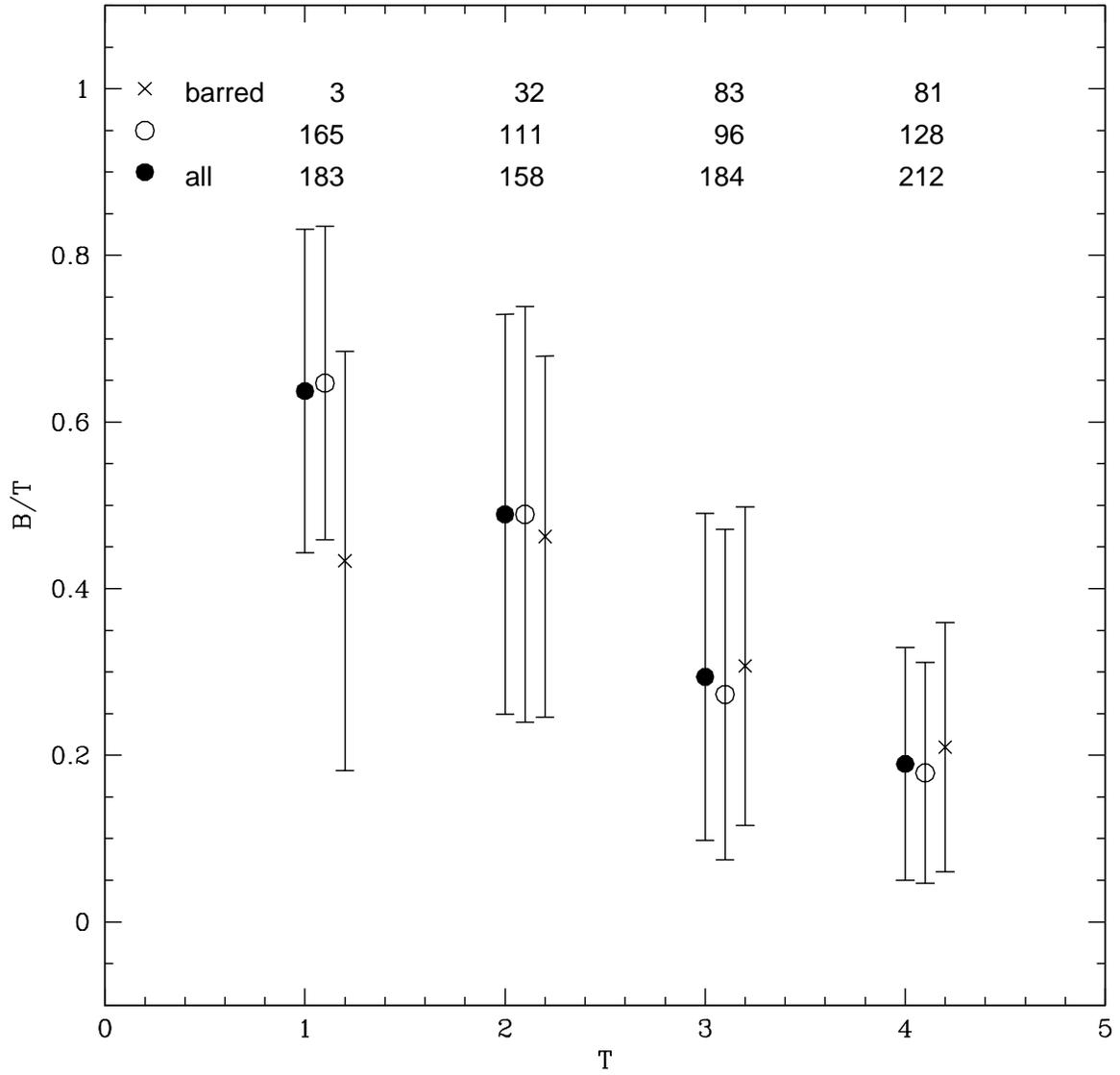}
\caption{Bulge to total luminosity ratio $B/T$ as a function of
morphological types of disk galaxies. The crosses, open
circles, and filled circles represent barred, non-barred, and
total sample, respectively. Barred/non-barred classification was
uncertain for 38 galaxies, which are not included in the plot.
\label{fig2}}
\end{figure}

\begin{figure}
\plotone{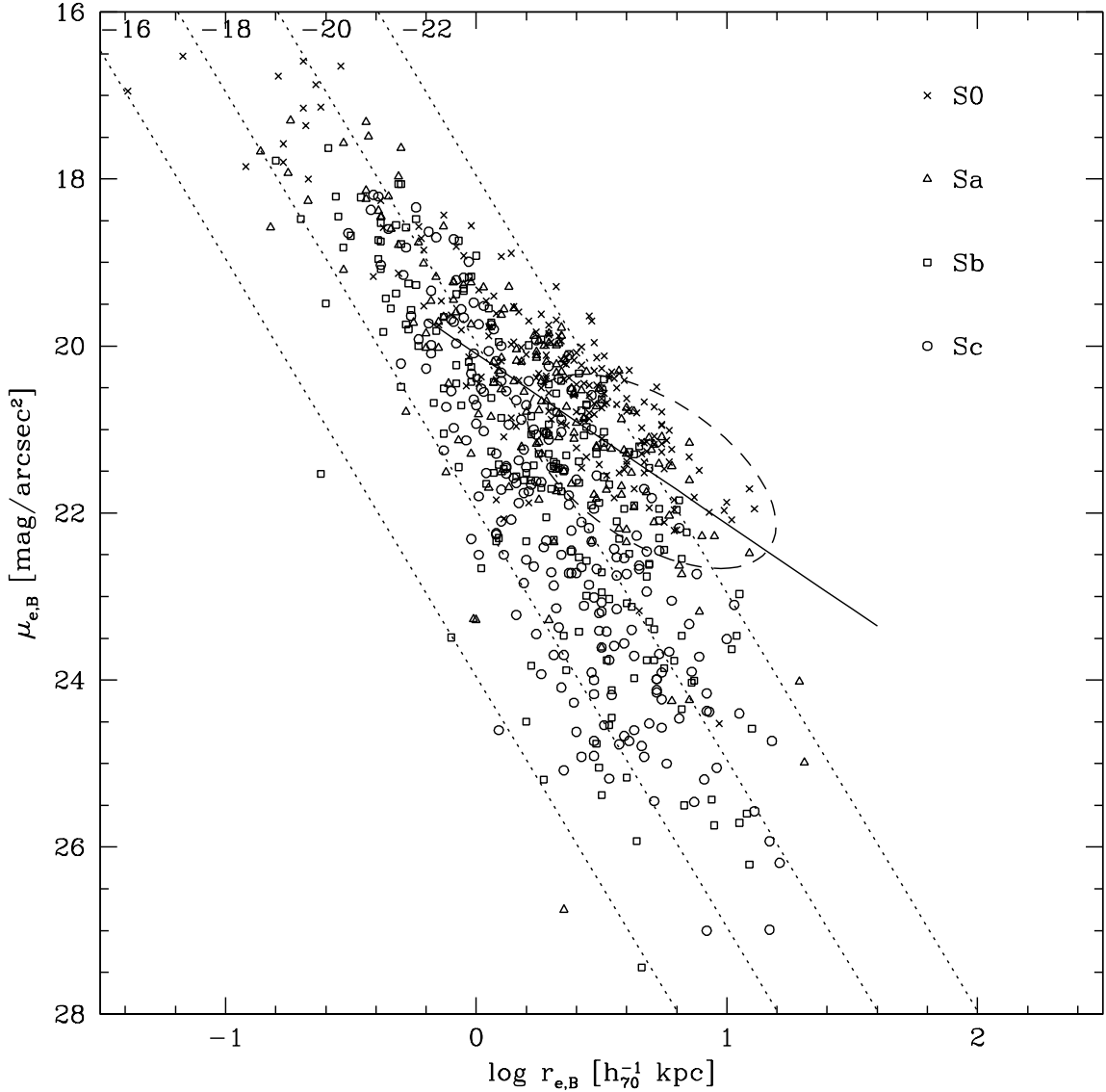}
\caption{Relation between the effective radius $r_{e,B}$ and
surface brightness $\mu_{e,B}$ at the effective radius for bulges.
Different symbols denote different morphologies of galaxies.
The solid line is the $\mu_{e,B}-r_{e,B}$ relation for
early-type galaxies from the SDSS 
(Bernardi et al. 2003), and the broken ellipse show the
area where most of the nearly 9000 SDSS early-type galaxies
are distributed. The dotted lines
are the relations that give rise to fixed bulge luminosities,
$M_{r,B}=-16, ...-22$ mag.
\label{fig3}}
\end{figure}

\begin{figure}
\plotone{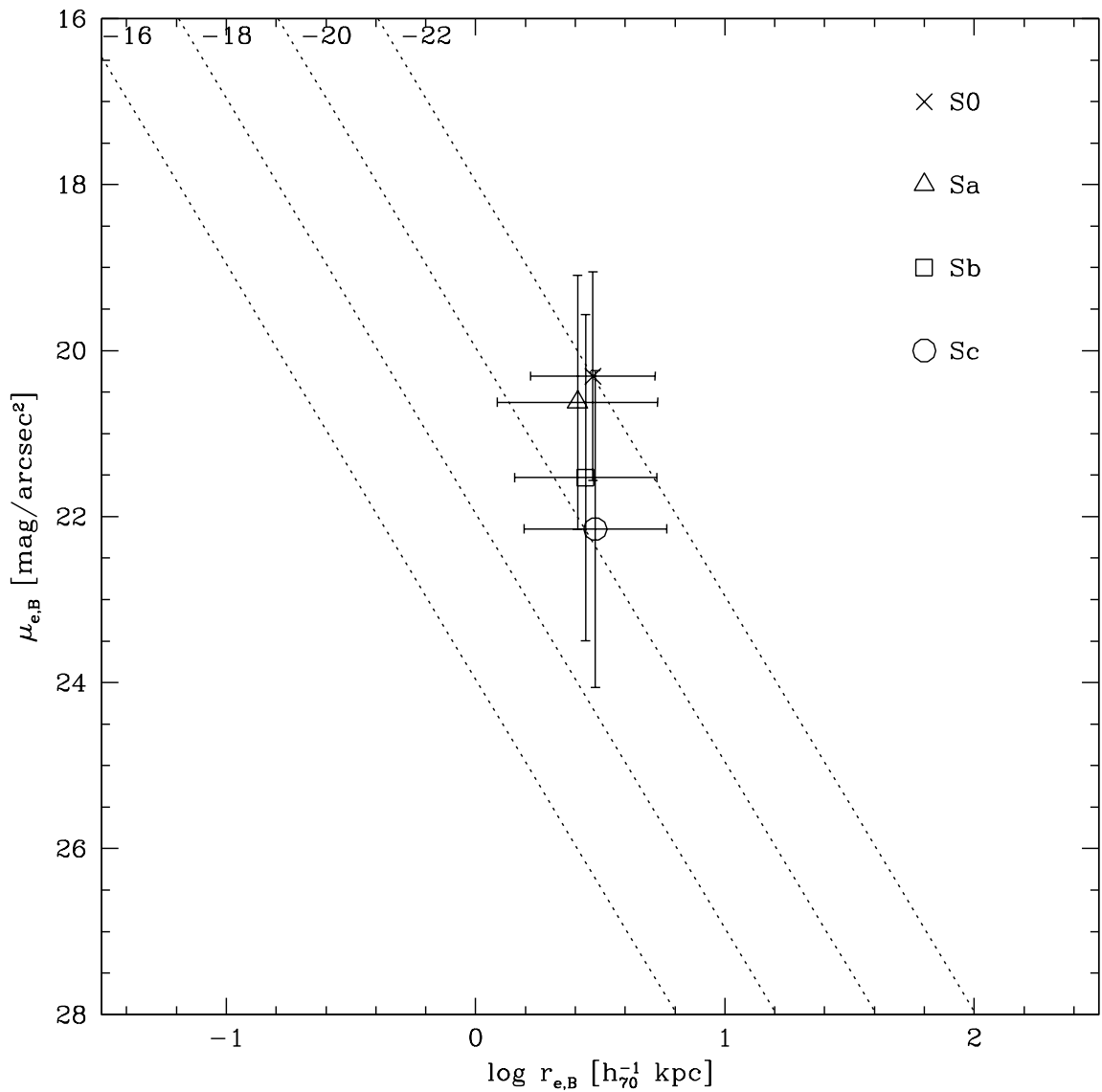}
\caption{Variation of the relation between the effective radius
$r_{e,B}$ and surface brightness $\mu_{e,B}$ at the effective 
radius for bulges, as morphologies changes.
The plot shows the mean and the dispersion of data shown in
Figure 3. Different symbols denote different morphologies of galaxies,
and dotted lines are the relations that give rise to fixed bulge 
luminosities, $M_{r,B}=-16, ...-22$ mag.
\label{fig4}}
\end{figure}

\begin{figure}
\plotone{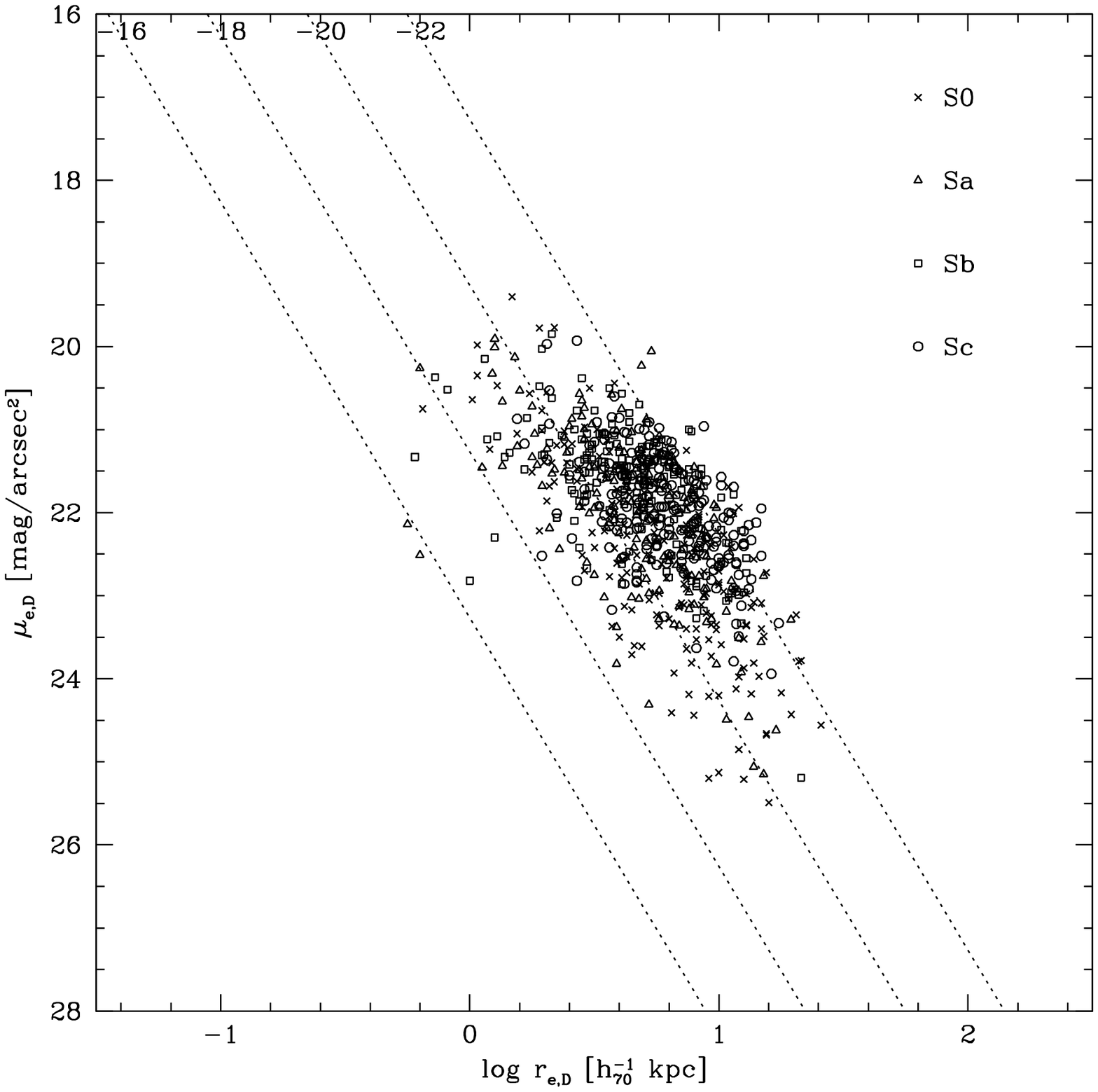}
\caption{Same as Figure 3, but for disks, i.e.,
the effective radius $r_{e,D}$ vs.
surface brightness $\mu_{e,D}$ at the effective radius.
The dotted lines
are the relations that give rise to fixed disk luminosities,
$M_{r,D}=-16, ...-22$ mag.
\label{fig5}}
\end{figure}

\begin{figure}
\plotone{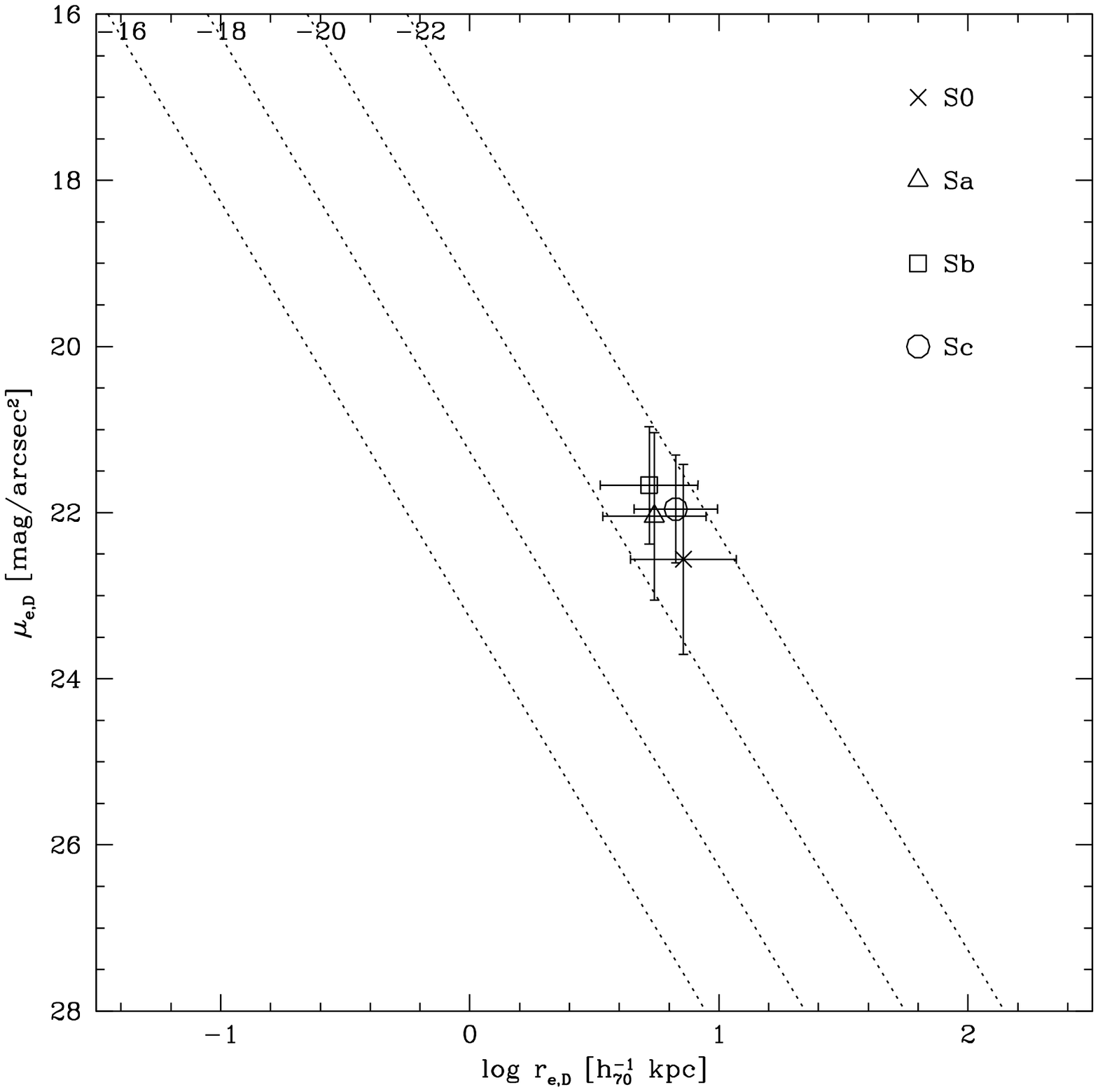}
\caption{Same as Figure 4,
but for disks, i.e.,
the effective radius $r_{e,D}$ vs.
surface brightness $\mu_{e,D}$ at the effective radius.
The dotted lines
are the relations that give rise to fixed disk luminosities,
$M_{r,D}=-16, ...-22$ mag.
\label{fig6}}
\end{figure}

\begin{figure}
\plotone{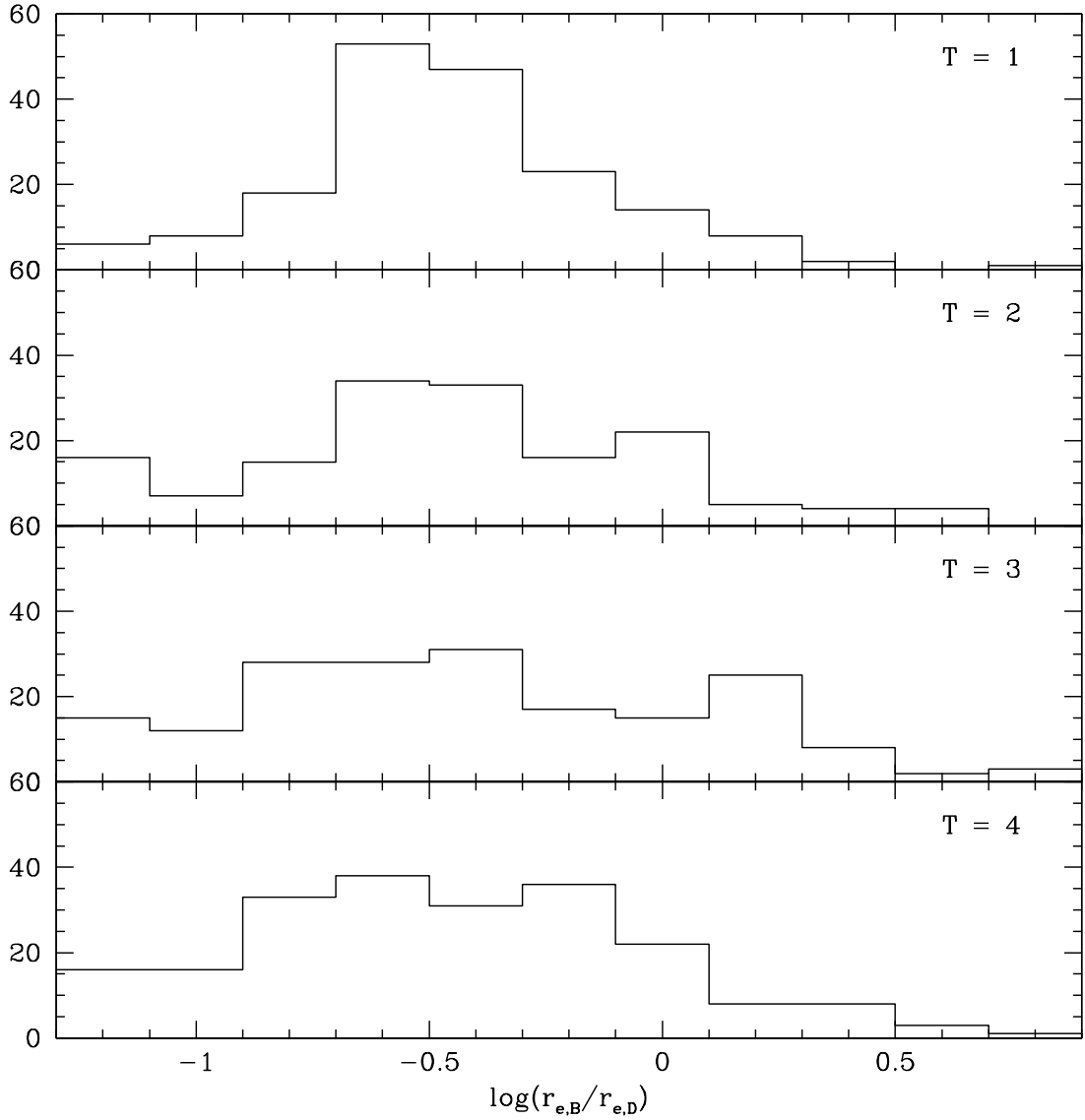}
\caption{Distribution of the ratio of scale lengths of bulges and disks
for the four classes of morphologies. There are no data points beyond
the
scale displayed.
\label{fig7}}
\end{figure}

\begin{figure}
\plotone{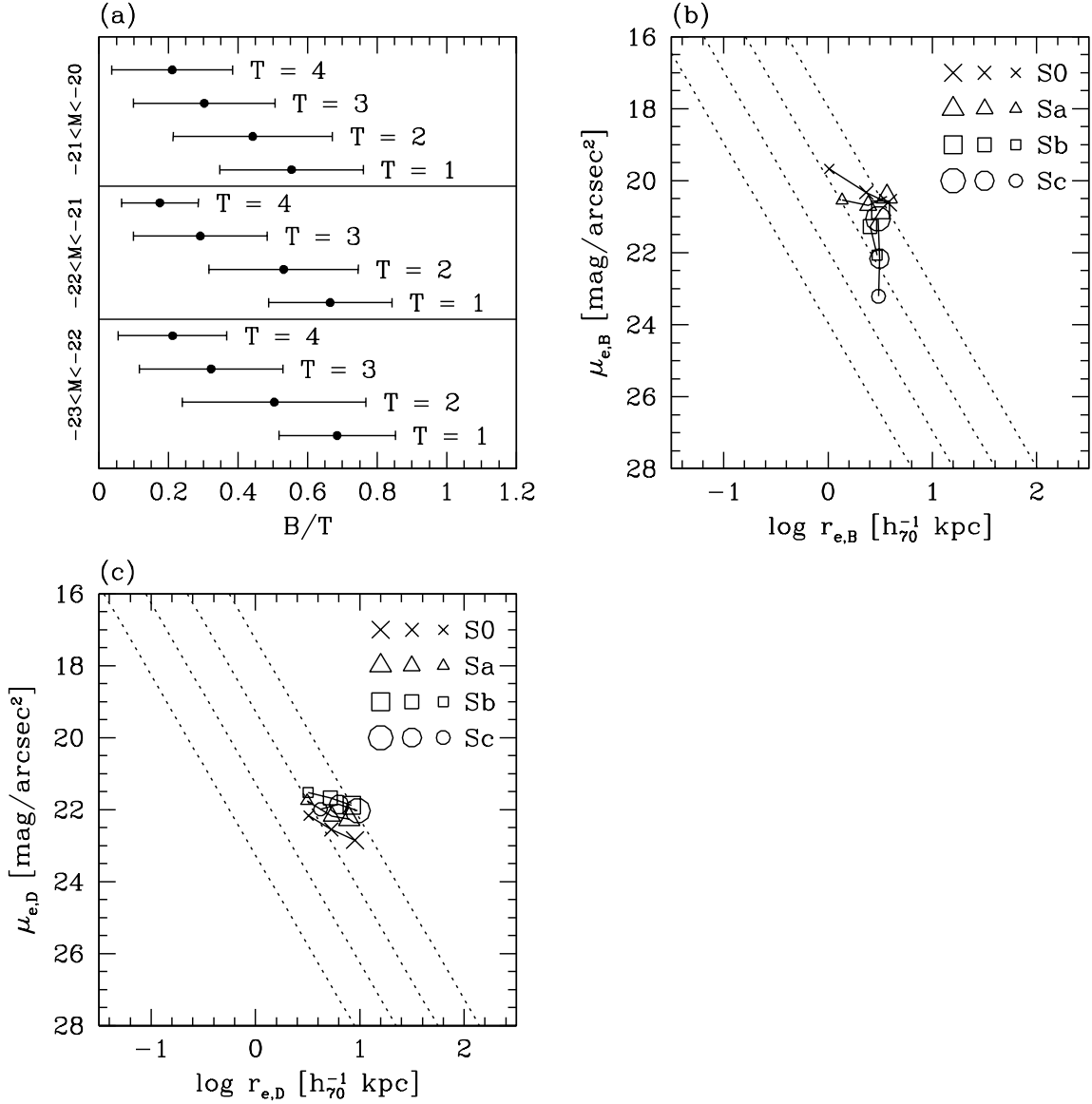}
\caption{(a) Bulge to total luminosity ratios for the four morphology
types for the three luminosity groups. The error bars show dispersions,
as for Figure 2. (b) the effective radius $r_{e,B}$ vs. surface 
brightness $\mu_{e,B}$ for bulges for the three luminosity
groups (denoted by the size of symbols). The large size symbols stand for
galaxies with $-23<M<-22$, middle size symbols for $-22<M<-21$ and small
size symbols for $-21<M<-20$. The dotted lines are the relations that give 
rise to fixed bulge luminosities, $M_{r,B}=-16, -18, -20, -22$ mag
from left to right.
(c) the same as (b) but for disks.
\label{fig8}}
\end{figure}

\begin{figure}
\plottwo{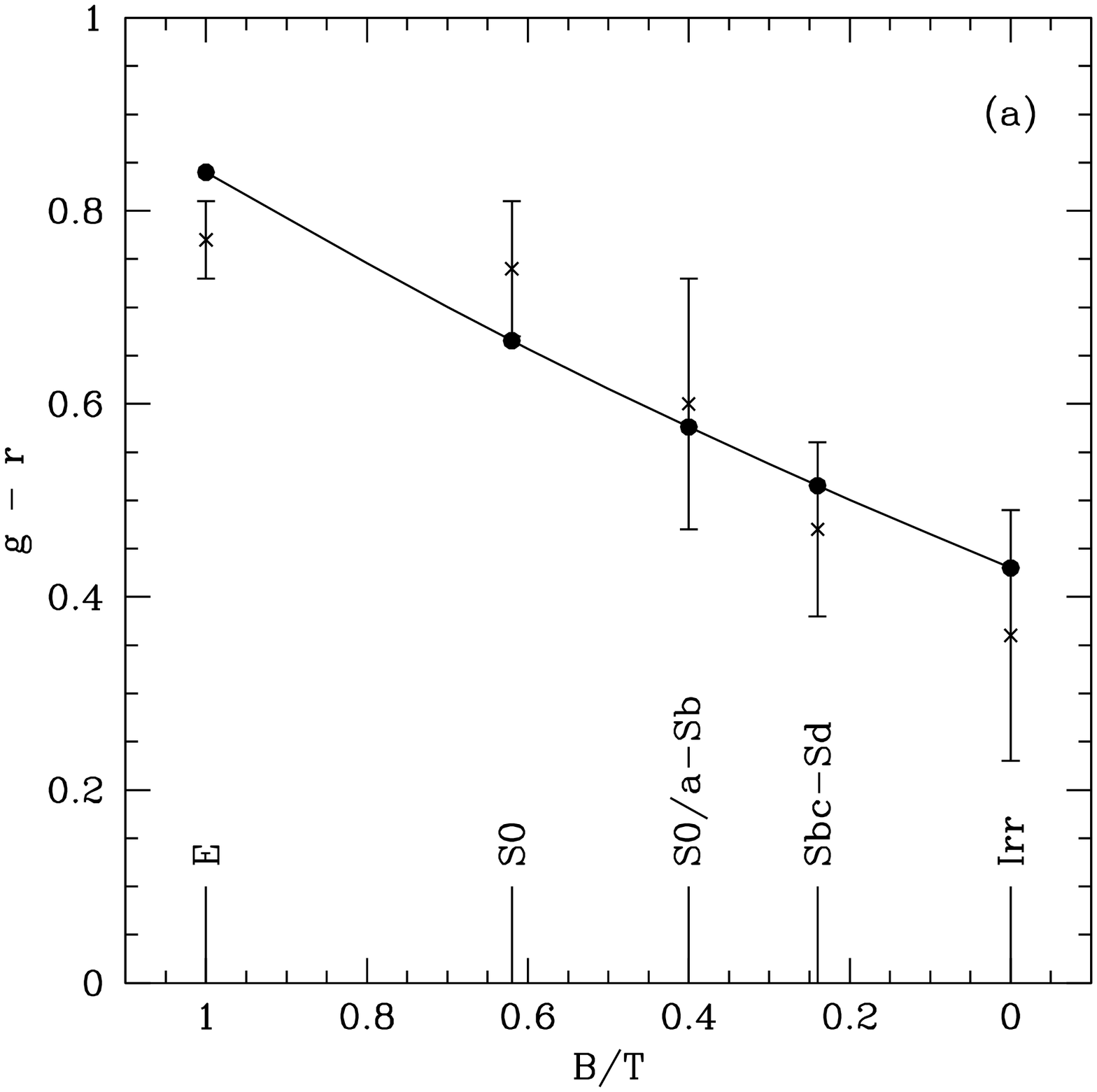} {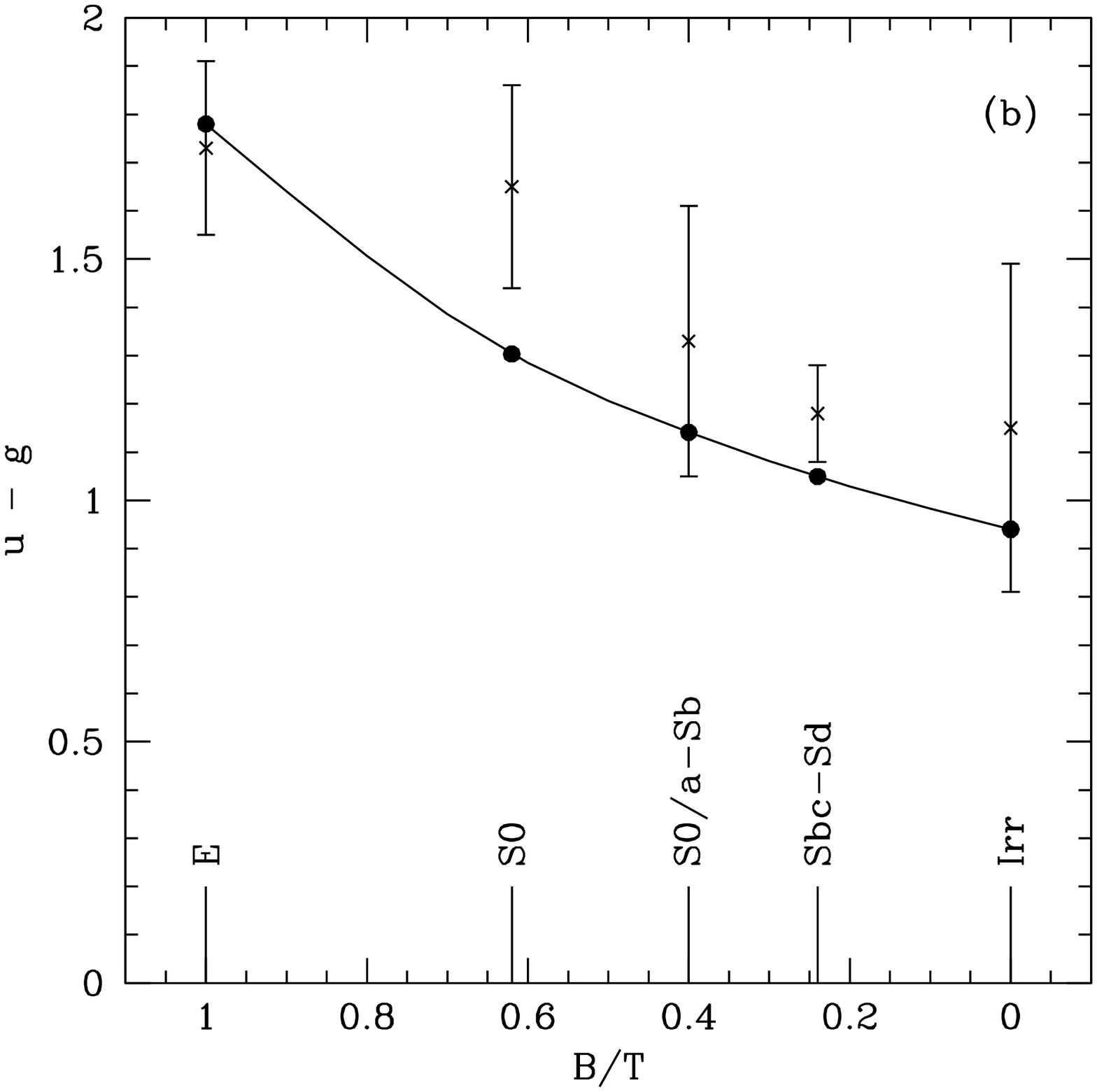}
\caption{$g-r$ color (left) and $u-g$ color (right) of galaxies in
various morphological types. The curves are prediction based
on the bulge disk decomposition given in this paper
with the aid of the
stellar population synthesis with a delayed exponential
star formation model. The data with error
bars (variance) are statistics from morphologically classified sample
of SDSS galaxies.}  
\label{fig9}
\end{figure}

\end{document}